\documentclass[conference]{IEEEtran}
\IEEEoverridecommandlockouts
\usepackage{cite}
\usepackage{amsmath,amssymb,amsfonts}
\usepackage{algorithmic}
\usepackage{graphicx}
\usepackage{textcomp}
\usepackage{xcolor}
\usepackage{float}
\usepackage{enumitem}
\usepackage{hyperref} 
\def\BibTeX{{\rm B\kern-.05em{\sc i\kern-.025em b}\kern-.08em
    T\kern-.1667em\lower.7ex\hbox{E}\kern-.125emX}}

\DeclareRobustCommand{\IEEEauthorrefmark}[1]{\smash{\textsuperscript{\footnotesize #1}}}

\begin{document}

\title{A Novel Multivariate Bi-LSTM model for Short-Term Equity Price Forecasting}

\author{\IEEEauthorblockN{Omkar Oak\IEEEauthorrefmark{1}, Rukmini Nazre\IEEEauthorrefmark{1}, Rujuta Budke\IEEEauthorrefmark{1}, Yogita Mahatekar\IEEEauthorrefmark{2}}
\IEEEauthorblockA{\IEEEauthorrefmark{1}Under Graduate Student, Department of Computer Science and Engineering, COEP Technological University}
\IEEEauthorblockA{\IEEEauthorrefmark{2}Assistant Professor, Department of Mathematics, COEP Technological University}
\IEEEauthorblockA{\IEEEauthorrefmark{1}omkarsoak@gmail.com, nazrerukmini@gmail.com, rujutabudke@gmail.com}
\IEEEauthorblockA{\IEEEauthorrefmark{2}yvs.maths@coeptech.ac.in}
}

\maketitle

\begin{abstract}
Prediction models are crucial in the stock market as they aid in forecasting future prices and trends, enabling investors to make informed decisions and manage risks more effectively. In the Indian stock market, where volatility is often high, accurate predictions can provide a significant edge in capitalizing on market movements. While various models like regression and Artificial Neural Networks (ANNs) have been explored for this purpose, studies have shown that Long Short-Term Memory networks (LSTMs) are the most effective. This is because they can capture complex temporal dependencies present in financial data. This paper presents a Bidirectional Multivariate LSTM model designed to predict short-term stock prices of Indian companies in the NIFTY 100 across four major sectors- ICICI Bank, NTPC, Ambuja Cement and Wipro.
The study utilizes eight years of hourly historical data, from 2015 to 2022, to perform a comprehensive analysis of the proposed methods. Both Univariate LSTM and Univariate Bidirectional LSTM models were evaluated based on R² score, RMSE, MSE, MAE, and MAPE. To improve predictive accuracy, the analysis was extended to multivariate data. Additionally, 12 technical indicators, having high correlation values with the close price(greater than 0.99) including EMA5, SMA5, TRIMA5, KAMA10 and the Bollinger Bands were selected as variables to further optimize the prediction models. The proposed Bidirectional Multivariate LSTM model, when applied to a dataset containing these indicators, achieved an exceptionally high average R² score of 99.4779\% across the four stocks, which is 3.9833\% higher than that of the Unidirectional Multivariate LSTM without technical indicators. The proposed model has an average RMSE of 0.0103955, an average MAE of 0.007485 and an average MAPE of 1.1635\%. This highlights the model's exceptional forecasting accuracy and emphasizes its potential to improve short-term trading strategies.
\end{abstract}


\section{Introduction}
The stock market is a critical component of the financial system, and its significant price fluctuations can have far-reaching effects on the broader economy \cite{1}. Accurate predictions of stock prices are essential for preventing market crashes, enhancing market management, and fostering financial stability. However, forecasting stock prices presents challenges due to the market's volatility, non-linearity, and the complex nature of financial data. While traditional methods have focused on historical data analysis to predict future trends, recent advancements in machine learning, particularly Long Short-Term Memory (LSTM) networks, have shown promise in improving prediction accuracy by capturing long-term patterns in time-series data \cite{2, 3, 4}.
This study utilizes both Univariate and Multivariate LSTM models to forecast short-term stock prices, with a focus on stocks from the National Stock Exchange of India (NSE). Given the high volatility of the stock market, effective prediction mechanisms are crucial for managing risks and making informed investment decisions, especially in India where over 190 million people actively trade stocks \cite{8}. By analyzing historical market data, this study aims to improve resource management and trend forecasting.
Hourly data is used to capture daily market fluctuations while considering computational complexity, with the goal of identifying trading opportunities and enhancing short-term price trend analysis. The research focuses on predicting short-term closing prices for stocks from four major sectors—Finance, Energy, Industrial, and Information Technology (IT)—which are significant contributors to India’s economy. The selected stocks are among the top contributors to the NIFTY 100 index, making them key indicators of market performance. Additionally, backtesting is employed to evaluate the effectiveness of trading strategies and predictive models using historical data. This process helps researchers and investors assess the potential success of their approaches before real-world application, thereby reducing financial risk and improving decision-making accuracy.

\section{Literature Review}
Stock price prediction has evolved from traditional statistical methods like ARIMA (AutoRegressive Integrated Moving Average) and GARCH (Generalized Autoregressive Conditional Heteroskedasticity) to Machine Learning (ML) techniques and, more recently, deep learning approaches to enhance forecasting accuracy \cite{9}. Statistical methods, foundational in econometrics, capture patterns in historical data but may struggle with non-linear relationships and complex patterns in modern markets. For instance, Menon et al. (2016) successfully used ARIMA and GARCH for bulk price forecasting \cite{10}, while Shakhla et al. (2018) applied multiple linear regression with success \cite{11}.\\
ML models, including Naive Bayes(NV),Random Forests (RF) and Support Vector Machines (SVM), are able to overcome these limitations by leveraging their ability to handle large datasets and complex variable interactions \cite{12}. Patel et al.(2015) applied these algorithms to predict stock and stock index movements, demonstrating superior performance compared to traditional statistical approaches. However, despite their advantages, these methods often require extensive hyperparameter tuning, can be computationally intensive, and suffer from overfitting, especially when applied to volatile financial markets.\\
 ANNs were designed to learn intricate relationships through multiple hidden layers and activation functions and used by Guresen et al. for stock market index prediction. These models can handle large volumes of data with high dimensionality and uncover deep, non-linear dependencies that ML techniques might miss\cite{13}. However, a drawback of ANNs is that they cannot deal with temporal dependencies and sequential patterns.\\
Recurrent Neural Networks (RNNs), overcoming these temporal dependencies, handle sequential data by maintaining a form of memory through feedback connections. This enables them to process sequences more effectively than ANNs. Though RNNs can efficiently handle temporal data\cite{14}, they face issues like vanishing gradients, which affect their performance over long sequences.\\
Long Short-Term Memory (LSTM) networks are a major improvement for time series prediction, addressing the issues faced by RNNs. LSTMs include memory cells capable of retaining information over extended periods, lessening the vanishing gradient problem and allowing accurate capturing of long-term dependencies. Hochreiter and Schmidhuber (1997) first introduced LSTM networks, and emphasized that they could represent intricate temporal patterns\cite{4}. LSTM models possess short-term memory and can catch the longer term effects and predict financial data accurately\cite{15}. LSTMs outperform traditional econometric models in forecasting foreign exchange rates\cite{16}. Additionally, LSTMs can achieve better results than classical ML methods in predicting stock market movements\cite{17}. Taking these factors into consideration, this study focuses on LSTM as the primary architecture.\\
Unlike traditional LSTMs that process data in a single direction, bidirectional LSTMs analyze sequences in both forward and backward directions \cite{18}. This approach allows them to utilize information from both past and future contexts, offering a more comprehensive understanding of the data. This makes them especially efficient for stocks \cite{19}. Notably, Han et al. implemented a Bidirectional LSTM model, with a mean squared error of 0.00020 \cite{20}. We propose to use technical indicators to further improve upon the model performance.\\
Incorporation of technical indicators, like moving averages and relative strength index, leads to further improvement in the accuracy of prediction models. Alsubaie et al. in their 2019 study, found that using at least 10 technical indicators will lead to a higher prediction accuracy on stock price data\cite{21}.\\
In this study, we present a Bidirectional Multivariate LSTM model for short term trade prediction, which incorporates the use of technical indicators.

\section{Methodology}
\subsection{Dataset}\label{AA}
The stocks used for the study are ICICI, NTPC, Ambuja Cement and Wipro. Historical prices of a total of eight years from Jan 2015 to Feb 2022 were extracted from the Yahoo Finance website \cite{22}.  The data samples were of 5-minute intervals, which were converted to hourly data during preprocessing. Each final stock dataset thereby had 10,862 entries. We then used TA-lib \cite{23}, an open source library for Technical Analysis of financial data to generate the following 50 technical indicators. The input and output pairs were then created using a window size of 24. A train, test and validation split of 70\%, 15\% and 15\% respectively was used. The training dataset had 7586 values, test 1626 and validation 1626.

\begin{figure}[h!]
\centering
\includegraphics[scale=0.3]
{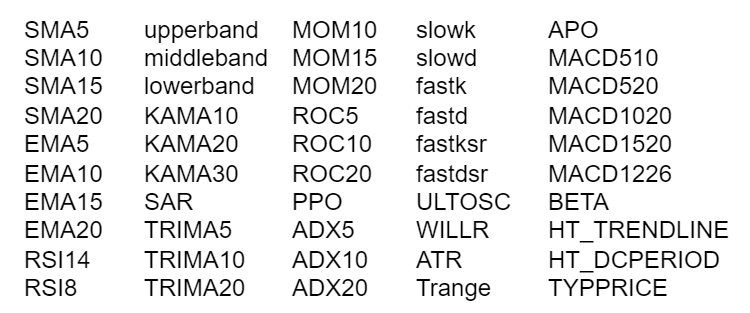}
    \caption{List of All Technical Indicators}
    \label{fig1}
\end{figure}

A brief explanation of Open High Low Close and Volume (OHLCV) metrics:
\begin{itemize}
\item Open Price: The open price represents the initial price at which a stock begins trading during a specified period.
\item High Price: It is the highest price reached by a stock within a given time frame.
\item Low Price: It is the lowest price recorded by the stock during a specific time window.
\item Closing Price: It indicates the value of a stock at the end of a particular time frame.
\item Volume: Volume is the number of shares traded (both sold and bought) in all within a selected period, typically on a daily basis.
\end{itemize}

\subsection{Data Preprocessing}
The data was first converted to hourly intervals. 
We decided to analyze the importance of the generated indicators and select only those which would improve the prediction models’ performance.
\subsubsection{Selection of technical indicators}
We calculated the average correlation of each technical indicator with the close price over all four stocks, and selected the 12 most highly correlated values (greater than 0.99), including open, high, close, low and volume. 
\begin{equation*}
r = \frac{\sum(x_i-\bar{x})(y_i-\bar{y})}{\sqrt{\sum(x_i-\bar{x})^2\sum(y_i-\bar{y})^2}}
\end{equation*}

The correlation formula as given below was used. Where $y_i$ and $x_i$ represent individual data points, whereas $\bar{y}$ and $\bar{x}$ are the variables' means. The selected indicators were SMA5, EMA5, TRIMA5, KAMA10, Lowerband, Middleband and Upperband.

\subsubsection{Normalization}
\begin{figure}[h!]
\centering
\includegraphics[scale=0.32]
{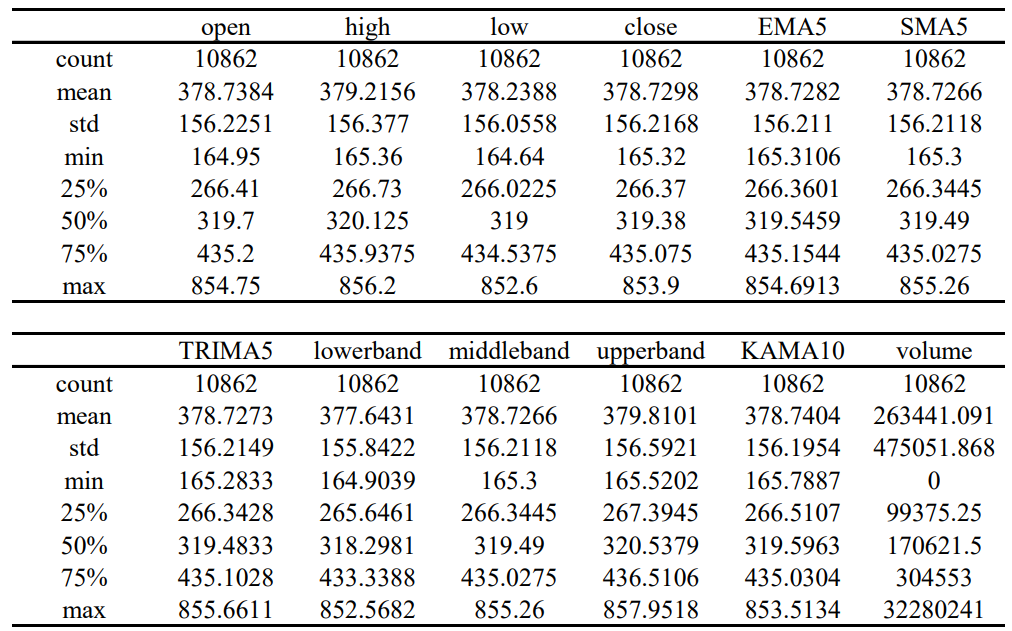}
    \caption{Data Description for ICICIBANK}
    \label{fig2}
\end{figure}

After selecting technical indicators, we performed an exploratory data analysis of each stock dataset. For example, Figure \ref{fig2} gives the distribution for ICICIBANK.
Taking into consideration the large range of the data and its volatility - extreme values, negative values and so on, we decided to normalize each column, ensuring all values lie in the range [0,1], so as to ensure better performance of the neural networks. 

\subsection{Technical Indicators}
The Simple Moving Average (SMA) is used to calculate average price over a given period, smoothing out the data to find trends.
The Exponential Moving Average (EMA) gives greater weightage to recent prices, thus being more sensitive to any new information.
The Triangular Moving Average (TRIMA) smooths price data with greater emphasis on the middle period of the calculation range \cite{book1}. 
Bollinger Bands are made up of one middle band, SMA, and two outer bands which reflect price volatility. The lower and upper bands are set at a distance of typically two standard deviations with respect to the middle band. They help identify overbought or oversold conditions and volatility.
\begin{table}[h!]
\caption{Technical Indicators and Their Formulas}
\resizebox{\columnwidth}{!}{%
\begin{tabular}{cc} \hline
\textbf{Indicator} & \textbf{Formula} \\ \hline
SMA5               & $\frac{1}{5}\sum_{i=1}^{5} C_i$ \\  \vspace{5px}
EMA5               & $C_t \times \frac{k}{5+1} + EMA_{t-1} \times \left(1-\frac{k}{5+1}\right)$ \\ \vspace{5px}
KAMA10             & $KAMA_t = KAMA_{t-1} + \left(\text{SC}_t \times (C_t - KAMA_{t-1})\right)$ \\ \vspace{5px}
TRIMA5             & $\text{TRIMA5} = \text{SMA}(\text{SMA}(C_i,5),5)$ \\ \vspace{5px}
Lower Band         & $MB - (D \times \sigma_i)$ \\ \vspace{5px}
Middle Band        & $\frac{1}{20}\sum_{i=1}^{20} C_i$ \\ \vspace{5px}
Upper Band         & $MB + (D \times \sigma_i)$ \\ \hline
\end{tabular}%
}
\label{techincal indicators}
\end{table}
Kaufman’s Adaptive Moving Average (KAMA) adjusts based on market volatility, filtering out noise by adapting its sensitivity. 
Table \ref{techincal indicators} contains formulae for the selected indicators, where $C_i$ stands for close price, $SC$ is the smoothing constant in KAMA, and $D$ is the deviation multiplier in Bollinger Bands.

\subsection{Prediction models}
\subsubsection{LSTM Cell Architecture}
The cell state (C) in an LSTM acts as long-term memory, preserving information across time steps for learning long sequences, while the hidden state (h) represents the immediate output, summarizing the cell state for predictions. LSTM cells have three gates: the input gate, which controls how much new input is added to the cell state; the forget gate, which decides what to discard from the cell state; and the output gate, which filters the cell state's information to influence the hidden state. The cell state is updated by combining the old state with a new memory candidate, guided by the input and forget gates.
\subsubsection{Unidirectional LSTM}
Unidirectional LSTM models process data in one direction, typically from past to future, relying solely on historical information. Univariate LSTM focuses on forecasting a single time series, such as stock prices, making it suitable for simpler tasks where only past data is necessary. 

\begin{figure}[h!]
\centering
\includegraphics[scale=0.2]
{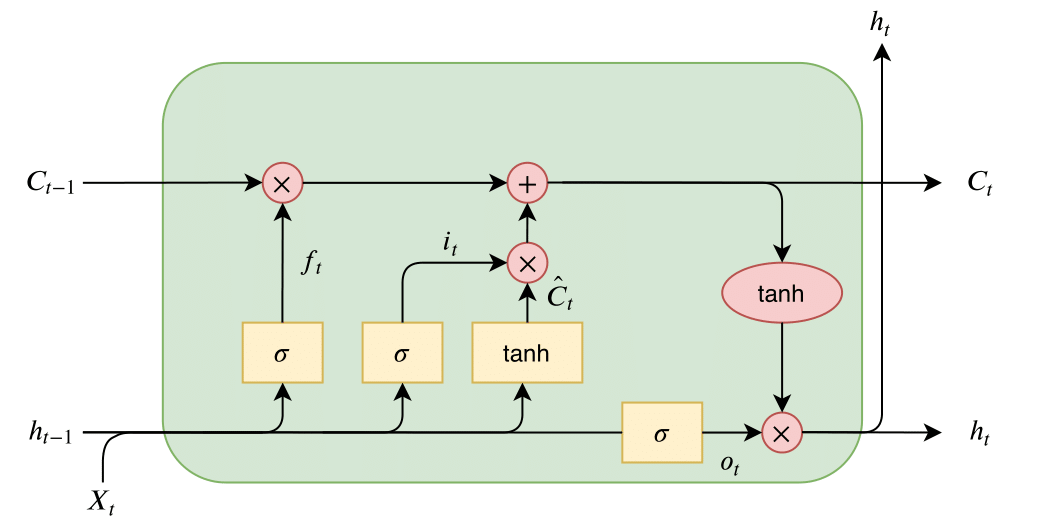}
    \caption{Unidirectional LSTM Cell Architecture}
    \label{fig3}
\end{figure}

In contrast, Multivariate LSTM handles multiple time series or features, like stock prices and trading volumes, simultaneously. This approach captures interactions between variables, leading to improved forecasting accuracy and a deeper understanding of how various factors impact the target variable. By processing data sequentially, unidirectional LSTMs are effective for scenarios where future context is not required for accurate predictions.

\subsubsection{Bidirectional LSTM}
Introduced by Graves et al. in 2005 \cite{18}, Bidirectional LSTM (BiLSTM) models enhance classical LSTM architectures by processing data in the forward direction as well as the backward direction. This architecture enables the capture of dependencies from past contexts and from future contexts relative to each time step, offering an increasingly comprehensive grasp of temporal patterns. 

\begin{figure}[h!]
\centering
\includegraphics[scale=0.2]
{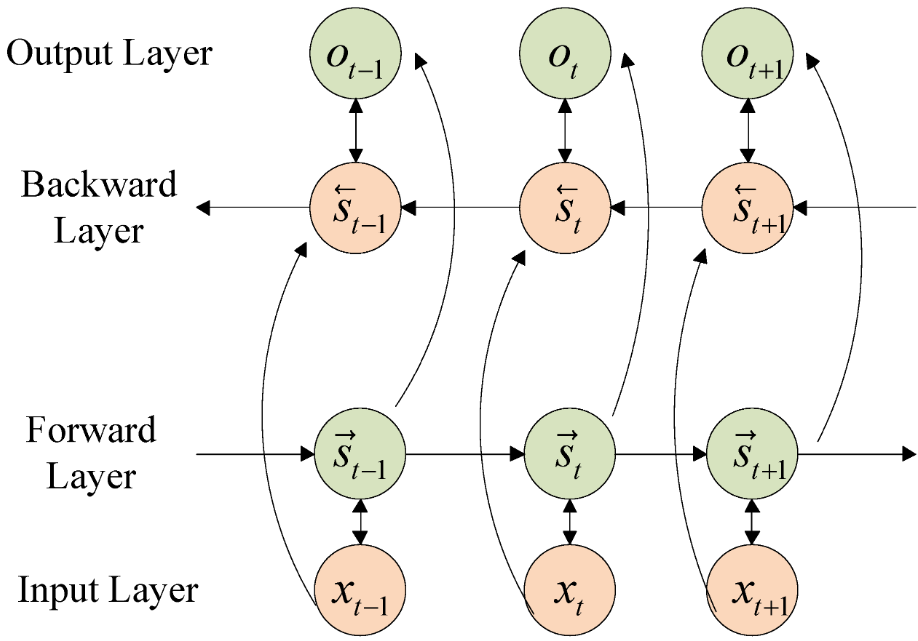}
    \caption{Bidirectional LSTM architecture}
    \label{fig4}
\end{figure}

Our proposal focuses on the Bidirectional Multivariate LSTM, which extends this concept to multiple time series or features, such as stock prices and trading volumes. By processing these features in both directions, the model can capture complex interactions and correlations between variables from both past and future perspectives. This processing enables the model to integrate a richer set of information, providing a more nuanced understanding of how various factors influence each other and the target variable. The enhanced ability to model intricate relationships and dependencies improves prediction accuracy and offers deeper insights into the dynamics of the data.

\subsection{Prediction Approaches}
\begin{enumerate}[label=(\alph*)]
\item \textbf{Univariate -- Close Value approach:} Here, we used only the closing price data to predict future close prices.
\item \textbf{Multivariate -- OHLCV Approach:} Here, we used the open, high, low, close and volume features to predict close prices.
\item \textbf{Multivariate -- Technical Indicators Approach:} Here, in addition to OHLCV, we used technical indicators as well.  
\end{enumerate}

\begin{figure}[h!]
\centering
\includegraphics[scale=0.15]
{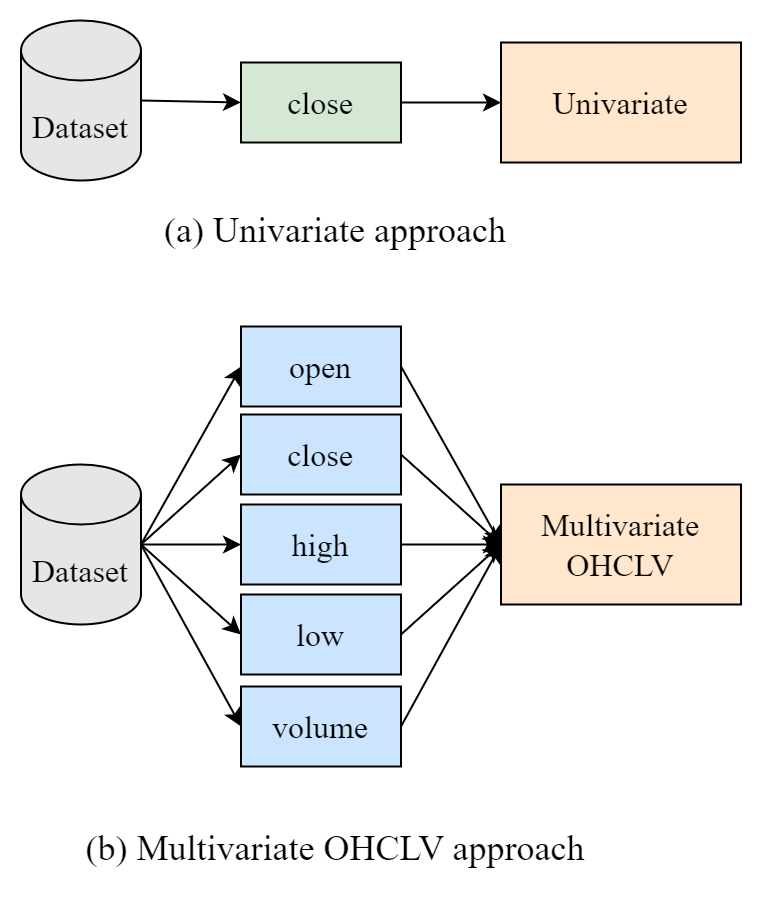}
    \caption{Traditional approaches}
    \label{fig5}
\end{figure}

\begin{figure}[h!]
\centering
\includegraphics[scale=0.12]
{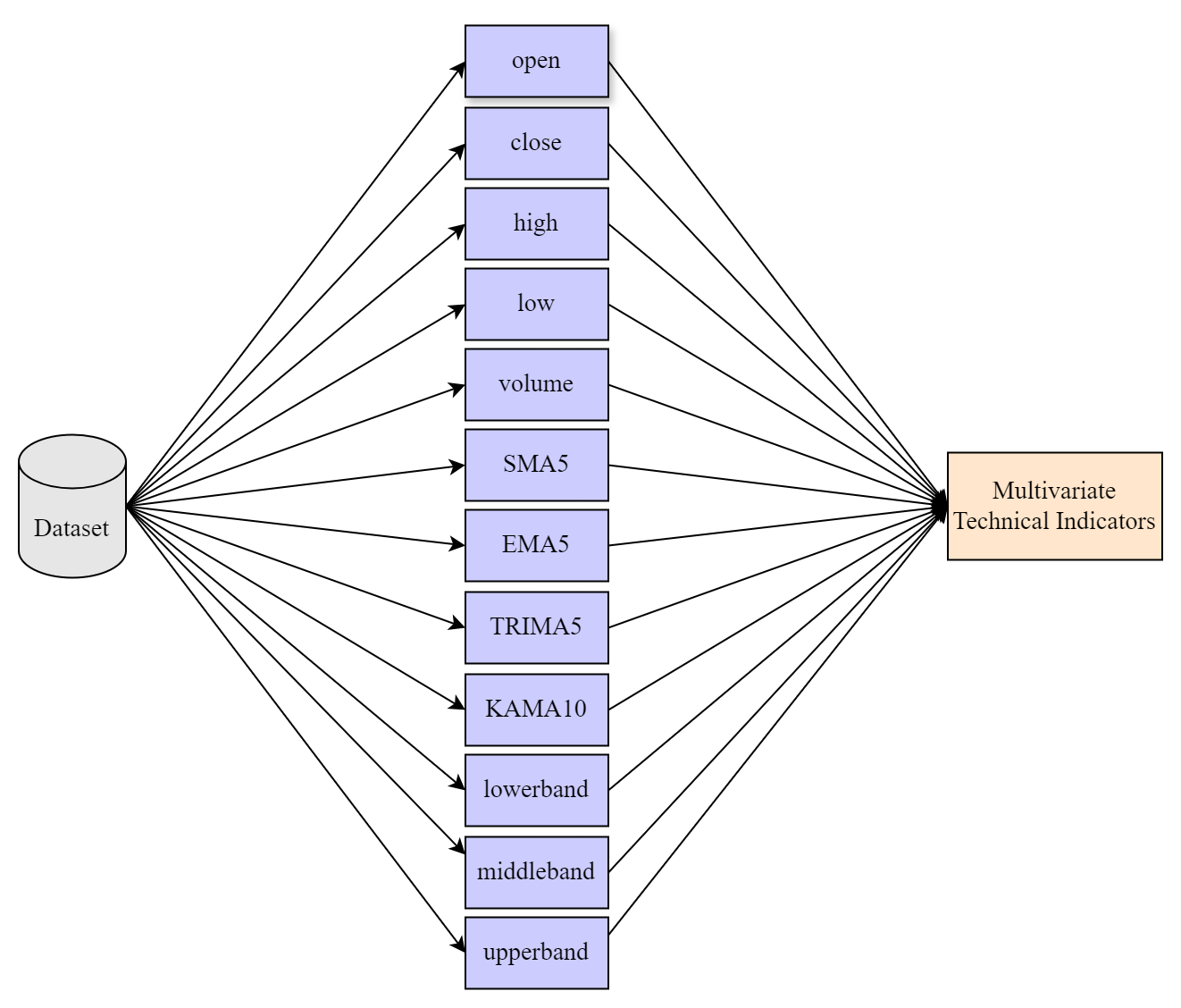}
    \caption{Proposed Multivariate Technical Indicators approach}
    \label{fig6}
\end{figure}

Figure \ref {fig5} shows the selected input variables for the univariate and multivariate OHCLV approaches, and Figure \ref{fig6} shows the proposed multivariate approach with technical indicators.

\section{Implementation and Results}
\subsection{Evaluation Measures}
The following standard measures were used to derive a robust performance evaluation for the considered models, taking into consideration the nature of the data and the intended task:
\begin{enumerate}
\item $R^2$ Score: Also called the coefficient of determination, it is a statistical measure used in ML for evaluating regression models. By evaluating the percentage of the dependent variable's variance that the independent variables account for, it determines how well a model fits the data. 
\begin{equation*}
    R^2 = 1 - \frac{SS_{R}}{SS_{T}}
\end{equation*}
Mathematically, it is calculated by the comparison of the Sum of Squared Residuals (SSR) with the Total Sum of Squares (SST). 

\item Mean Absolute Error (MAE): 
\begin{equation*}
    \text{MAE} = \frac{1}{n} \sum_{i=1}^{n} \left| Y_i - \hat{Y}_i \right|
\end{equation*}
MAE calculates the average magnitude of errors between the actual and predicted values without considering whether they are positive or negative. It is a straightforward measure of model accuracy, and reflects the average error in the same units as those of the data.

\item Mean Absolute Percentage Error (MAPE): 
MAPE measures the average percentage error between the predicted and actual values. It is often used to express forecast accuracy as a percentage, and is easy to interpret. However, it can be sensitive to extremely small actual values, which then leads to very high percentage errors.
\begin{equation*}
    \text{MAPE} = \frac{1}{n} \sum_{i=1}^{n} \left| \frac{Y_i - \hat{Y}_i}{Y_i} \right| \times 100\%
\end{equation*}

\item Root Mean Squared Error (RMSE): 
RMSE is calculated as the square root of the MSE, bringing the error measure back to the original units of the data. It is widely used because it penalizes large errors more than MAE, while still being interpretable with respect to units.
\begin{equation*}
    \text{RMSE} = \sqrt{\frac{1}{n} \sum_{i=1}^{n} (Y_i - \hat{Y}_i)^2}
\end{equation*}

\end{enumerate}

\subsection{Sliding Window Approach}
In this study, we employ a sliding window approach to forecast the next time step's equity price using historical price data. The sliding window technique is a very frequently used method in time series analysis, especially while dealing with sequential data using RNNs such as LSTM networks.
\subsubsection{Window Size Selection}
Selecting a window size, denoted as $W$, is critical for capturing temporal dependencies in the data. For our experiments, we selected a window size of $W=24$ hours, corresponding to $24$ sequential hourly stock prices. This window size was determined based on preliminary experiments and domain-specific considerations, ensuring that the model captures daily price trends while avoiding overfitting to shorter-term fluctuations.
\subsubsection{Formulation}
Given a time series of stock prices $S={s_1,s_2,\dots,s_T}$, where $s_t$ is the stock price at time $t$, the sliding window approach creates overlapping sequences of length $W$ to be used as input features. For each prediction, the model utilizes the past $W$ prices to forecast the price at the next time step $t+1$. \\
Mathematically, for a given time $t$, the input-output pair can be defined as:
\begin{equation*}
X_t={s_{t-W+1},s_{t-W+2},\dots,s_t}, \;\; y_t=s_{t+1}  
\end{equation*}
Where $X_t$ represents the input sequence consisting of the stock prices from $t-W+1$ to $t$, and $y_t$ represents the predicted stock price at time $t+1$.

\subsection{Model Training and Results}
All models were run on each of the 4 stocks, in identical hardware and software environments. Using Google Colab's T4 GPU, the models were trained for 10 epochs, using an initial learning rate of 0.001. The Adam optimizer was used, and the loss function was mean squared error. Each of the models had early stopping with a patience factor of 5.

\begin{table}[h!]
\caption{ICICIBANK Results}
\resizebox{\columnwidth}{!}{%
\begin{tabular}{lccccc} \hline
    Model & $R^2$ & MAE & RMSE & MAPE \\ \hline
    Univariate LSTM &               0.9826 & 0.010906 & 0.014003 & 1.4012 \\ 
    Univariate bi-LSTM &            0.9893 & 0.007936 & 0.010972 & 1.0712 \\ 
    Multivariate OHCLV LSTM &       0.9241 & 0.023189 & 0.029181 & 2.8648 \\ 
    Multivariate OHCLV bi-LSTM &    0.9922 & 0.006668 & 0.009364 & 0.8863 \\ 
    Multivariate LSTM &             0.9879 & 0.008702 & \textbf{0.001163} & 1.1295 \\ 
    Multivariate bi-LSTM &          \textbf{0.9926} & \textbf{0.006532} & 0.009096 & \textbf{0.8639} \\ \hline
\end{tabular}%
}
\label{table2}
\end{table}

\begin{table}[h!]
\caption{WIPRO Results}
\resizebox{\columnwidth}{!}{%
\begin{tabular}{lccccc} \hline
    Model & $R^2$ & MAE & RMSE & MAPE \\ \hline
    Univariate LSTM &               0.9935 & 0.009571 & 0.012951 & 1.3396 \\ 
    Univariate bi-LSTM &            0.9885 & 0.013873 & 0.017319 & 1.8516 \\ 
    Multivariate OHCLV LSTM &       0.9216 & 0.037438 & 0.045227 & 4.7221 \\ 
    Multivariate OHCLV bi-LSTM &    0.9870 & 0.015423 & 0.018392 & 2.1552 \\ 
    Multivariate LSTM &             0.9732 & 0.021153 & 0.026422 & 2.6542 \\ 
    Multivariate bi-LSTM &          \textbf{0.9961} & \textbf{0.007396} & \textbf{0.010097} & \textbf{1.0236} \\ \hline
\end{tabular}%
}
\label{table3}
\end{table}

\begin{figure}[h!]
\centering
\includegraphics[scale=0.07]
{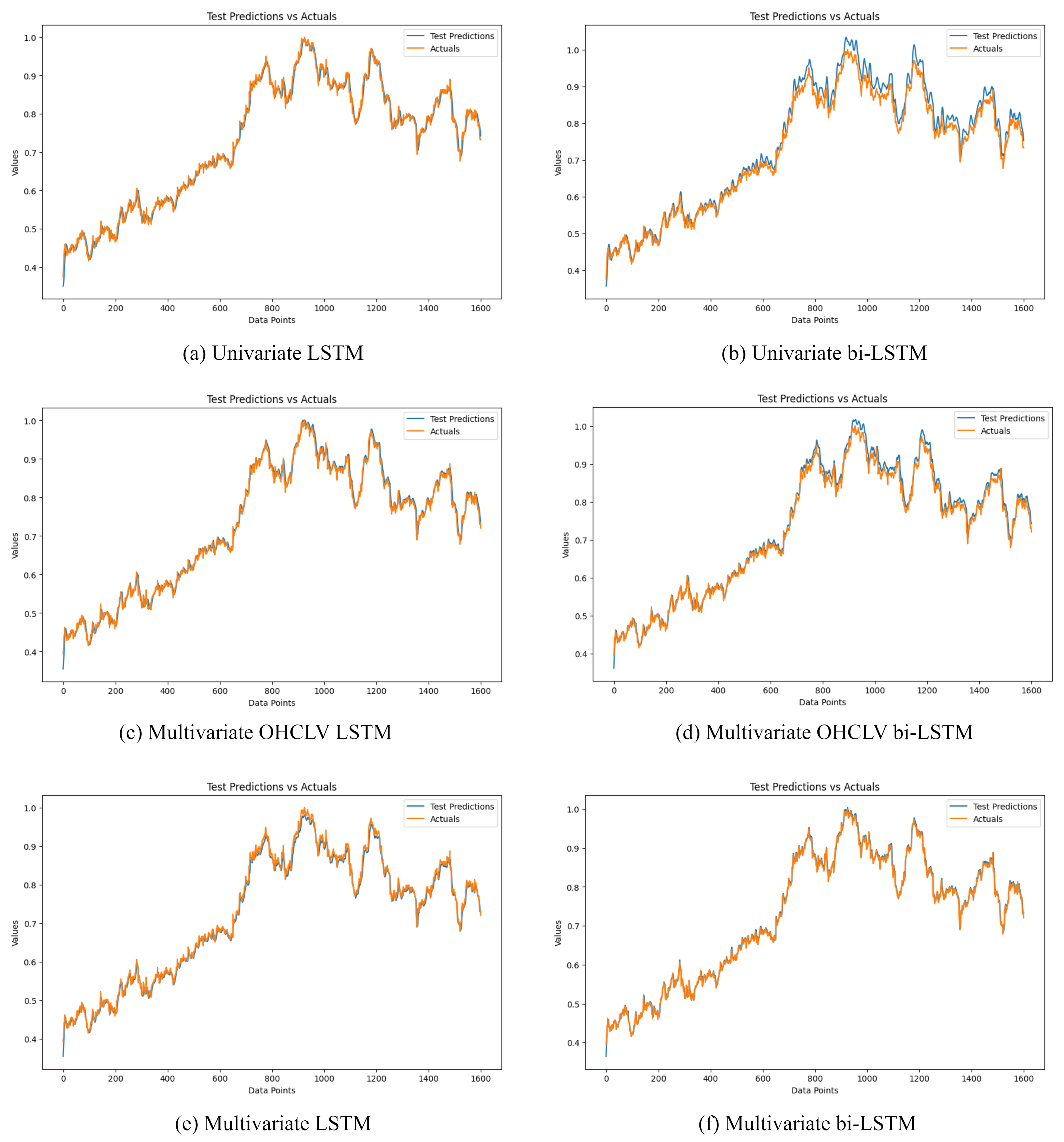}
    \caption{Prediction Plots of AMBUJACEM}  
    \label{fig11}
\end{figure}

As can be seen from the results tables \ref{table2}, \ref{table3}, \ref{table4}, and \ref{table5}, the proposed Bidirectional LSTM with technical indicators shows the best performance on all stocks, achieving an $R^2$ score of 99.67\% on Ambuja Cement and 99.61\% on Wipro, with MAE values of 0.006584 and 0.007396, respectively. This can also be seen from the model prediction plots on the Ambuja Cement dataset in Figure \ref{fig11}.

\begin{table}[h!]
\caption{NTPC Results}
\resizebox{\columnwidth}{!}{%
\begin{tabular}{lccccc} \hline
    Model & $R^2$ & MAE & RMSE & MAPE \\ \hline
    Univariate LSTM &                   0.9492 & 0.031262 & 0.037446 & 5.4748 \\ 
    Univariate bi-LSTM &                0.9891 & 0.012964 & 0.017321 & 2.5136 \\ 
    Multivariate OHCLV LSTM &           0.9803 & 0.017054 & 0.023464 & 3.1943 \\ 
    Multivariate OHCLV bi-LSTM &        0.9899 & 0.013085 & 0.016812 & 2.4457 \\ 
    Multivariate LSTM &                 0.9910 & 0.012214 & 0.015862 & 2.2621 \\ 
    Multivariate bi-LSTM &              \textbf{0.9937} & \textbf{0.009430} & \textbf{0.013310} & \textbf{1.8296} \\ \hline
\end{tabular}%
}
\label{table4}
\end{table}

\begin{table}[h!]
\caption{AMBUJACEM Results}
\resizebox{\columnwidth}{!}{%
\begin{tabular}{lccccc} \hline
    Model & $R^2$ & MAE & RMSE & MAPE \\ \hline
    Univariate LSTM &                   0.9929 & 0.009896 & 0.013360 & 1.4271 \\ 
    Univariate bi-LSTM &                0.9770 & 0.019744 & 0.024054 & 2.6028 \\ 
    Multivariate OHCLV LSTM &           0.9938 & 0.009361 & 0.012525 & 1.3354 \\ 
    Multivariate OHCLV bi-LSTM &        0.9899 & 0.012644 & 0.015975 & 1.6891 \\ 
    Multivariate LSTM &                 0.9942 & 0.009257 & 0.012055 & 1.2908 \\ 
    Multivariate bi-LSTM &              \textbf{0.9967} & \textbf{0.006584} & \textbf{0.009079} & \textbf{0.9369} \\ \hline
\end{tabular}%
}
\label{table5}
\end{table}

In contrast, the Univariate LSTM shows the weakest performance, with higher MAE and RMSE values, such as 0.010906 and 0.014003 for ICICI, and lower $R^2$ scores. Adding Open High Low and Volume statistics improved the prediction performance, as compared to univariate prediction, and led to higher accuracy.

The results indicate that incorporating technical indicators as input variables enhances model performance by providing additional features that capture market dynamics like momentum, volatility, and trends. These indicators help the model learn more complex patterns and reduce overfitting. The use of Bidirectional LSTMs further improves performance by processing the sequence data in backward as well as forward directions, allowing the model to capture dependencies from both future and past time steps. This bidirectional approach improves the model's ability to learn robust temporal patterns and address long-term dependencies with greater accuracy than traditional LSTMs. 

\section{Conclusion and Future Scope}
This paper aims to develop a novel technique for forecasting stock price trends, specifically tailored for short-term trades using hourly data from ICICI, NTPC, Ambuja Cement, and Wipro over an eight-year period (2015–2022). We employed Unidirectional and Bidirectional LSTM models for both Univariate and Bivariate data. Three research approaches were explored: the first used a basic univariate forecast model with stock close price data, the second included additional features such as open, high, low, and volume, and the third incorporated 12 features, including technical indicators that closely follow the trends of the close price.\\
The analysis revealed that the proposed Bidirectional LSTM model outperformed Unidirectional LSTMs, demonstrating a comprehensive understanding of price movements. Furthermore, the inclusion of technical indicators significantly improved the proposed model's performance across all sectors of the Indian National Stock Exchange, demonstrating it's robustness and adaptability to diverse market conditions. \\
The proposed model achieved the highest $R^2$ score of 99.6736\% on Ambuja Cement, underscoring the value of integrating technical indicators with advanced LSTM architectures. This makes it ideal for predicting stock movements in real time, leading to more profitable and secure investments across all major sectors of the Indian National Stock Exchange.\\
The proposed model offers potential for future improvements and wider applications. One approach is to extend it for long-term stock price predictions by using longer window sizes to capture broader market trends. Expanding the model to other major stock exchanges, like NYSE, Nikkei, and LSE, could validate its effectiveness across different market conditions. Additionally, incorporating macroeconomic indicators, social media sentiment, and global news could enhance its predictive power, leading to a more comprehensive approach to stock price forecasting.

\end{document}